# Probing the Invisible Universe: The Case for Far-IR/Submillimeter Interferometry

D. Leisawitz, T. Armstrong, D. Benford, A. Blain, K. Borne W. Danchi, N. Evans, J. Gardner, D. Gezari, M. Harwit, A. Kashlinsky, W. Langer, C. Lawrence, P. Lawson, D. Lester, J. Mather, S. H. Moseley, L. Mundy, G. Rieke, S. Rinehart, M. Shao, R. Silverberg, D. Spergel, J. Staguhn, M. Swain, W. Traub, S. Unwin, E. Wright, and H. Yorke

*Abstract*

The question "How did we get here and what will the future bring?" captures the human imagination and the attention of the National Academy of Science's Astronomy and Astrophysics Survey Commitee (AASC). Fulfillment of this "fundamental goal" requires astronomers to have sensitive, high angular and spectral resolution observations in the far-infrared/submillimeter (far-IR/sub-mm) spectral region.  With half the luminosity of the universe and vital information about galaxy, star and planet formation, observations in this spectral region require capabilities similar to those currently available or planned at shorter wavelengths.  In this paper we summarize the scientific motivation, some mission concepts and technology requirements for far-IR/sub-mm space interferometers that can be developed in the 2010-2020 timeframe.

1. Science goals

The Decade Report posed a number of "theory challenges," two of the most compelling of which are that astrophysicists should strive to: (a) *develop an "integrated theory of the formation and evolution of [cosmic] structure";* and (b) *"develop models of star and planet formation, concentrating on the long-term dynamical co-evolution of disks, infalling interstellar material, and outflowing winds and jets."* (Decade Report, p. 106)

Rieke et al. (2002; hereafter the "SAFIR white paper") explain the vital role that will be played by future far-IR/sub-mm observations in confronting these challenges and the need for a 10-m class Single Aperture Far Infrared Observatory (SAFIR). SAFIR will represent a factor $10^5$ gain in astronomical capability relative to the next-generation missions SIRTF and Herschel, yet it will have the visual acuity of Galileo's telescope. An additional hundred-fold increase in angular resolution can be achieved with interferometry after SAFIR and within the NASA Roadmap time horizon. In this section we explore the science potential of sub-arcsecond resolution in the far-IR/sub-mm, picking up where the SAFIR white paper leaves off. In particular, we don't bother to explain why the far-IR spectrum (line and continuum radiation) is rich in information content, as doing so would only restate facts already eloquently presented in the SAFIR white paper.

1.1 The heritage and destiny of cosmic structure

After we locate in space and time the first generations of stars, galactic bulges, galactic disks, and galaxy clusters, we will want to relate these early structures to the "seeds" of structure seen in the cosmic microwave background fluctuations and learn how they formed. We will need measurements that show us how the cosmic structures changed over time to the present day. We will want to lift the veil of dust that conceals galactic nuclei, including our own, from view at visible wavelengths. How did the Milky Way form, and why is there a black hole at its center? What happens to the interstellar medium when galaxies collide, and how does a starburst work?

Did bulges form first and disks form later, or did disks merge to form bulges? What accounts for the diversity of galaxy types? How might the universe and its constituents look when it is twice or ten times its current age?

It will take a telescope much bigger than 10 m to see structure in galaxies at redshift z ~ 1 or greater in the far-IR/sub-mm. These objects subtend angles of ~1 arcsec. SAFIR will measure far-IR spectra of huge numbers of high-z galaxies, and they will be analyzed statistically and with the aid of models and complementary NGST and ALMA observations. However, to study the astrophysics of distant galaxies it will be important to resolve them in the far-IR/sub-mm, where they emit half or more of their light (Trentham et al. 1999). As noted by Adelberger & Steidel (2000), high-z galaxies "are undeniably dusty…. Large corrections for dust extinction will be necessary in the interpretation of UV-selected surveys, and only IR observations can show whether the currently adopted corrections are valid or suggest alternatives if they are not." The far-IR spectrum tells us the amount of dust present, but says little about how the dust is distributed. The dust distribution, which will be seen directly when the galaxies are resolved in the far-IR/sub-mm, strongly influences the extinction (Calzetti 2001). The galaxy assembly process could be studied via high spatial resolution spectral line maps. For example, a $C^+$ 158 µm line map at $\lambda/\Delta\lambda \sim 10^4$ would provide vital information about the gas dynamics in merging and interacting systems and reveal the rotation speeds and velocity dispersions within and among galaxies and protogalaxy fragments.

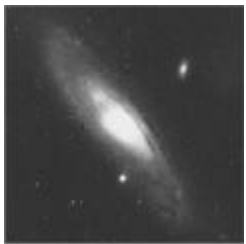

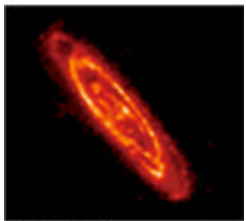

**Figure 1.** A far-IR/sub-mm interferometer that provides HST-class resolution would resolve as much detail in a galaxy at z = 10 as ISO did in M31. These images illustrate that it is impossible to deduce the far-IR appearance of a galaxy from an optical image. The far-IR image reveals the sites of star formation and the reservoir of interstellar matter available for new star formation (Haas et al. 1998).

As noted in the SAFIR white paper, far-IR continuum and line emissions are excellent indicators of the star formation rate and the physical conditions in star forming molecular clouds. At 10 Mpc, the distance of a nearby galaxy, a giant H II region subtends about 1 arcsec, and the typical spacing between neighboring regions is about 10 arcsec. SAFIR could be used to study individual sites of star formation spectroscopically. Later, with HST-class resolution, we could make similarly detailed observations of objects at much greater distances to learn how star formation works in protogalaxies and systems having very low heavy element abundance. With the same resolution we could study star forming regions in Virgo cluster galaxies at the linear scales sampled by past IR missions (IRAS and ISO) in the Milky Way. This would help us to understand the chemical and energetic effects of star formation on the interstellar and intergalactic medium and better interpret measurements of the high-z universe.

"The central regions of galaxies were likely heavily dust enshrouded during their formation epoch. Future far-IR observations can provide a window into this formation process and help determine the relationship between bulge formation and black hole formation." (Spergel, 2001) Black hole masses could be routinely measured with high spatial resolution spectral line mapping in the far-IR/sub-mm. High angular resolution submillimeter timing observations of the black hole at the Galactic center have the potential to enable a measurement of its spin (Melia et al. 2001). Such a measurement could substantially advance our understanding of the role played by supermassive black holes in galaxy formation and evolution (Elvis et al. 2002), and could yield new insight into fundamental physics, perhaps with cosmological implications.

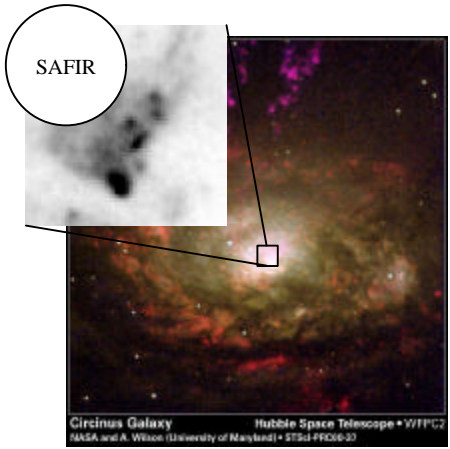

**Figure 2.** Multiple optical emission line sources were seen near the central active nucleus of Circinus with HST (Wilson et al. 2000). These sources would lie within a single SAFIR beam (circle). A far-IR/sub-mm interferometer could produce high-resolution images and spectral line maps and provide valuable information about the physical conditions, gas dynamics, and star formation in Active Galactic Nuclei, unequivocally testing AGN emission and orientation hypotheses.

High resolution is important for another, subtler reason. At wavelengths $\lambda > 200$ µm, the sensitivity of SAFIR will be confusion limited to ~10 µJy (Blain 2000); at these long wavelengths it could detect starburst galaxies out to $z > 5$, but $L_*$ galaxies only out to $z \sim 2$. However, an observatory with only three times better angular resolution than SAFIR would have a significantly lower confusion limit and could detect even a galaxy like the Milky Way out to $z \sim 10$. At far-IR/sub-mm wavelengths galaxies do not decrease in brightness with increasing redshift as $(1+z)^{-2}$, as one might expect, because an increasing portion of the emission is shifting *into* the observed wavelength band. At sub-mm wavelengths this so-called "negative K correction" compensates cosmological dimming out to $z \sim 10$. While a single aperture telescope larger than SAFIR may be possible, a factor of three seems very challenging. However, the nature of the far-IR sources is such that adequate sensitivity can be achieved with smaller apertures, and hence the spatial resolution can better be provided with interferometry. Thus, by beating confusion, far-IR/sub-mm interferometers could follow up on *all* the galaxies and proto-galaxies seen by HST, NGST, and ALMA. An important observational goal is to sample a representative volume of the high-z universe in the far-IR/sub-mm with HST-class angular resolution and spectral resolution sufficient to resolve the velocity structure in distant objects.

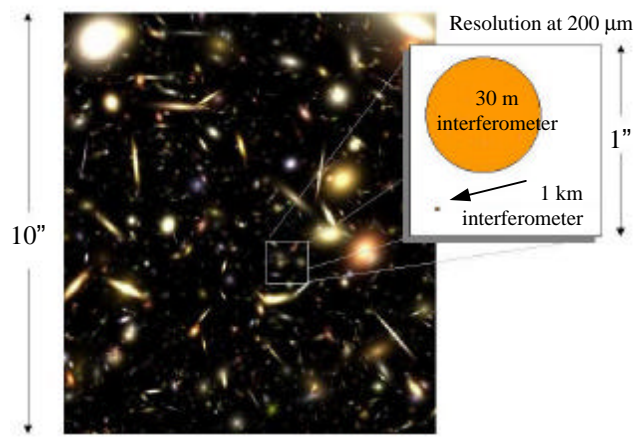

**Figure 3.** A far-IR/sub-mm interferometer with $10^{-20}$ W/m$^2$ sensitivity (equivalent bolometric magnitude 31.2) would slice through the Milky Way and between nearby galaxies to image galaxies and protogalactic objects out to $z \sim 10$. (Credit: A. Benson near-IR simulation for NGST)

It would be a great scientific achievement to image the pristine molecular hydrogen that must have allowed primordial gas clouds to cool, collapse, and give birth to the first generation of stars, before any heavier elements existed (Haiman et al. 1996). The most likely signature is a pair of H$_2$ cooling lines (rest wavelengths 17 and 28 µm) redshifted to $z > 10$ (Abel et al. 2002). SAFIR could detect this emission if it arises at $z < 10$, but its discovery may have to await a far-

IR/sub-mm interferometer if it comes from higher redshifts and is concentrated in discrete objects. An interferometer would resolve out confusing background emission and could have sufficient sensitivity to make the measurement.

1.2 The formation of small structures: stars, planets, and their inhabitants

How did the solar system and the Earth form? What are the various possible outcomes of the star and planet formation process, and how does the process work? How are the initial conditions in protostellar disks reflected in the properties of planetary systems? What chemical processes occur during star and planet formation?

Star and planet formation are parts of a single process that involves the movement of matter from envelopes extended over about 10,000 AU to disks on scales of 1 - 100 AU, and ultimately into stars and planets on much smaller scales (Evans 2001). The nearest protostellar objects are at 140 pc, where 1 AU subtends an angle of 7 mas. Future astrophysicists will need high spatial and spectral resolution measurements that reveal the bulk flows of material and the physical conditions (density, temperature, magnetic field strength, and chemical abundances) in dense molecular cores, protostars, protoplanetary systems, and debris disks (Evans 1999). The SAFIR white paper explains the essential need for far-IR continuum and spectral line measurements of these systems and the capability of SAFIR to resolve protostars down to the 100 AU scale.

With a far-IR/sub-mm interferometer we will be able to probe much smaller physical scales, particularly the scales relevant to studies of planet formation. Far-IR interferometric studies of circumstellar disks will reveal dust concentrations that represent the early stages of planet formation, and measurements of exozodiacal debris disks will show gaps and structures produced by resonances with already-existing planets (Ozernoy et al. 2000). By observing planetary systems in a wide range of evolutionary states and following individual systems over a period of years we could learn how protostellar material migrates and coalesces to form planets. Observations such as these would almost surely revolutionize our understanding of how the solar system formed. The rich far-IR line spectrum would be exploited to "follow the evolution of chemical abundances and locate reservoirs of biogenic materials" (Evans 2001). We will want to understand chemical evolution from molecular cores to planets in a unified way. The spectra of gas giant planets, which emit most of their light in the far-IR, could be measured with an interferometer.

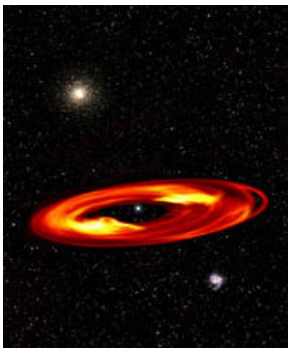

**Figure 4.** This artist's concept of the Vega debris disk illustrates the resonance features that could be studied with far-IR/sub-mm interferometers. At Vega's distance 1 AU subtends an angle of 128 mas. Studies of the structure of protoplanetary and debris disks will go a long way toward advancing our understanding of planet formation. (Credit: D. Wilner, M. Holman, P. Ho, and M. Kuchner; CfA Press release http://cfa-www.harvard.edu/newtop/previous/011802.html).

The next generation of far-IR and submillimeter observatories – SIRTF, SOFIA, and Herschel – will be very sensitive, but will have insufficient spatial resolution to achieve the observational goals outlined above. The key to making further progress will be to increase angular resolution by many orders of magnitude without sacrificing sensitivity or spectral resolution. About a decade from now ALMA will provide unprecedented spatial and spectral resolution at millimeter and submillimeter wavelengths, far out on the Rayleigh-Jeans tail, and probe regions with very high dust column densities. SAFIR will look where protostars are most luminous, in the far-IR, image them at ~100 AU scales for the first time, and chart the velocity structure to provide definitive evidence of envelope collapse. An additional factor of 10 - 100x improvement in spatial resolution will be needed to image protoplanetary and planetary debris disk structure in the spectral region where these objects emit the bulk of their energy.

2. Desired measurement capabilities

Table 1 summarizes the measurement capabilities needed to achieve the science goals outlined in sections 1.1 and 1.2 and shows that the desired capabilities are similar for the two applications. Table 2 summarizes the capabilities of the next-generation observatories NGST, SAFIR, and ALMA. SAFIR will pry open the door to the "invisible" far-IR universe and leave the astrophysics community desiring the next critical capability: better angular resolution. As can be seen by comparing the SAFIR column of Table 2 with Table 1 an improvement by two orders of magnitude in angular resolution is desired.

**Table 1**. Desired Measurement Capabilities for the Mid-IR to Millimeter Spectral Range

| Science goal | Formation and evolution of cosmic structure | Formation of stars and planetary systems |
|---|---|---|
| Sample targets | Hubble Deep Fields, gravitational lens sources, interacting galaxies | Nearest protostars, Orion prolyds, Vega, HH 30, and other disks |
| Wavelength range (peak emission) (μm) | 40 – 1000 | 30 – 300 |
| Angular resolution (mas) | 20 | 10 |
| Spectral resolution ($\lambda/\Delta\lambda$) | >$10^4$ | $3\times10^5$ |
| Point source sensitivity, $\nu S_\nu$ (W/m$^2$) | $10^{-20}$ | $10^{-20}$ |
| Field of view (arcmin) | 4 | 4 |

**Table 2**. Measurement Capabilities of Next-Generation Observatories

| Observatory | NGST | SAFIR (10 m) | ALMA |
|---|---|---|---|
| Wavelength range (μm) | 0.6 – 30 | 30 – 300 | 850 –10,000 plus windows at 350, 450 |
| Angular resolution (mas) | 50 at 2 μm | 2500 at 100 μm | 10 at 1 mm |
| Spectral resolution ($\lambda/\Delta\lambda$) | $10^3$ | $10^6$ | >$10^6$ |
| Point source sensitivity, $\nu S_\nu$ (W/m$^2$) | $10^{-21}$ – $10^{-20}$ | $10^{-20}$ | $10^{-19}$ at 1 mm |
| Field of view (arcmin) | 4 | 4 | 0.3 at 1 mm, bigger field with mosaicing |

3. Mission concepts

How will we satisfy the inevitable desire for detailed far-IR/sub-mm views of the high-redshift universe and protoplanetary disks? An interferometer with total aperture comparable to that of SAFIR (78 m$^2$) would have the desired sensitivity *and* could provide the desired angular resolution (Table 1). The resolution of an interferometer with maximum baseline $b_{max}$ is $\Delta\theta = 10$ mas ($\lambda/100$ μm)($b_{max}/1$ km)$^{-1}$. Thus, a 1 km maximum baseline is needed to provide the angular resolution ultimately desired in the far-IR/sub-mm. To obtain excellent image quality all spatial frequencies would have to be sampled in two dimensions; in other words, measurements would have to be made on many baselines $b < b_{max}$, and at many baseline position angles. This so-called "u-v plane" filling is accomplished with ground-based interferometers by deploying many apertures, allowing for array reconfiguration, and relying on Earth rotation. In space there is more freedom to move apertures to desired locations, so one can tailor the u-v coverage to the problem at hand. There is a substantial cost advantage to limiting the number of apertures, particularly because to achieve background-limited performance and the desired sensitivity, the mirrors would have to be very cold (~5 K). However, in space, where there is no atmosphere to distort the wavefront, 2 or 3 apertures would suffice. The preferred location for the interferometer is the Sun-Earth L2 point, as it is distant enough to help with cooling and pointing, yet near enough to handle a large data rate.

A Michelson interferometer, in which parallel beams are combined using a half-silvered mirror or the equivalent, offers several advantages. First, a relatively modest number of detectors would be required. In a conventional Michelson interferometer a single-pixel detector is needed for each baseline, or two such detectors can be used because there are two "output ports." Detector arrays would provide a multiplex advantage that could be used either to widen the field of view or improve signal-to-noise by spectrally dispersing. The field of view could be as large as 5 arcmin $(N_{pix}/100)$ $(\lambda/100\ \mu m)$ $(d/4\ m)^{-1}$, where d is the diameter of the individual aperture mirrors and $N_{pix}$ is the pixel count in one array dimension. A 100 x 100 pixel array would provide the desired field size. Second, a Michelson interferometer can be operated in "double Fourier" mode (Mariotti & Ridgway 1988), so it naturally provides high spectral as well as high spatial resolution. The spectral resolution $R = 10^4$ $(2\Delta/1\ m)$ $(\lambda/100\ \mu m)^{-1}$, where $\Delta$ is the length of the delay line stroke (i.e., $2\Delta$ is the optical delay), so a 0.5 m stroke would yield the desired spectral resolution in every spatial resolution element. A small additional optical delay would be needed to compensate for geometric delay associated with the off-axis angles in the wide field.

Mather et al. (1999) first suggested the possibility of a 1 km maximum baseline far-IR/sub-mm imaging and spectral interferometer space mission called SPECS (Submillimeter Probe of the Evolution of Cosmic Structure). The concept and a technology roadmap were further developed with science and engineering expertise provided through the February 1999 community workshop on "Submillimeter Space Astronomy in the Next Millennium" (http://space.gsfc.nasa.gov/astro/smm_workshop/). The concept of a science and technology pathfinder mission called SPIRIT (Space IR Interferometric Telescope) originated at the workshop. SPIRIT is much like SPECS, except that the interferometer would be built on a boom and have $b_{max}$ ~ 30 m ($\Delta\theta$ ~ 0.34 arcsec at 100 $\mu$m). SPECS, like the original concept for TPF, would use formation flying to maneuver the interferometer apertures. For more information on the SPIRIT and SPECS concepts see Leisawitz et al. (2000).

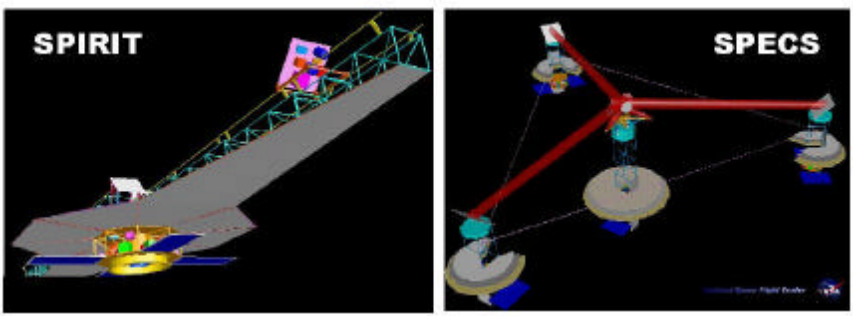

**Figure 5.** SPIRIT, a scientific and technology pathfinder for SPECS, could achieve the spatial resolution of a 60 m filled aperture telescope on a 30 m boom.

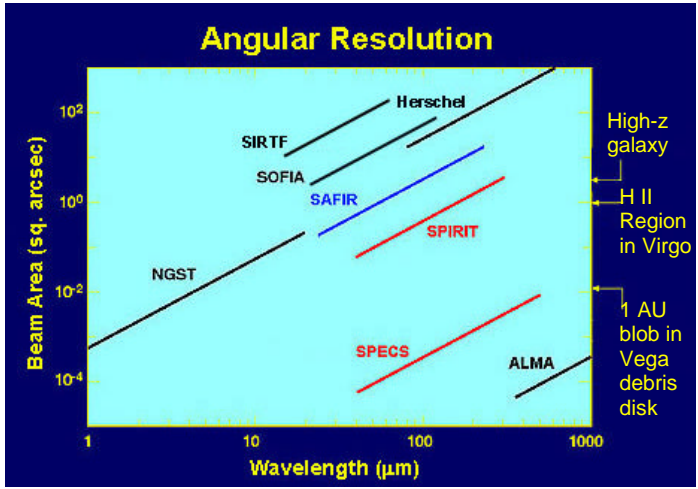

**Figure 6.** After SAFIR an additional hundred-fold improvement in angular resolution will be needed to achieve the science goals outlined in section 1 and attain resolution comparable to that of NGST and ALMA in the spectral regions that bracket the far-IR/sub-mm. The resolution gain could be accomplished in two steps, first with the pathfinder mission SPIRIT, and then with SPECS. N.B. A protoplanetary disk at 140 pc is about the same size as an H II region in the Virgo cluster.

4. Enabling technologies

*"The [Radio and Submillimeter Wave Astronomy] panel supports the recommendation of the Panel on Ultraviolet, Optical, and Infrared Astronomy from Space that NASA pursue technology development leading toward a far-infrared/submillimeter interferometer in space."* – Decade Report, Panel Reports, p. 170.

New technology will be needed in four areas: 1) detectors, 2) cooling, 3) optics and interferometry, and 4) large structures and formation flying. In this section we summarize the requirements in each area and cite possible solutions, then we conclude with a brief discussion of technology validation on space missions due to launch in the coming decade. More information on the enabling technologies for far-IR/sub-mm interferometry is given by Shao et al. (2000).

4.1 Detectors

The detector goal is to provide noise equivalent power less than $10^{-20}$ W Hz$^{-1/2}$ over the 40 – 850 µm wavelength range in a 100 x 100 pixel detector array, with low-power dissipation array readout electronics. This low noise level is a prerequisite for background-limited telescope performance. The ideal detector would count individual photons and provide some energy discrimination, which would enable more sensitive measurements. Among the encouraging recent developments in detector technology are superconducting transition edge sensor (TES) bolometers (Benford et al. 2002), SQUID multiplexers for array readout (Chervenak et al. 1999), and single quasi-particle counters built out of antenna-coupled superconducting tunnel junctions and Rf-single electron transistors (Schoelkopf et al. 1999).

4.2 Cooling

The cooling requirements for space-based far-IR/sub-mm interferometry are similar to those for a large single-aperture telescope like SAFIR. To take full advantage of the space environment, the mirrors will have to be very cold (~5 K) and the detectors even colder (<0.1 K). Active coolers will have to operate continuously and not cause significant vibrations of the optical components. The coolers should be light in weight. Cooling power will have to be distributed over large mirror surfaces. Thermal transport devices will likely have to be flexible and deployable. Large, deployable sunshades will be needed, and they will have to provide protection without seriously compromising sky visibility. Since several stages of cooling must be used to reach the required temperatures, the devices that operate in each temperature range must be able to interface with each other both mechanically and thermally. The Astro E-2 mission will use a three-stage cooling system for its X-ray microcalorimeters, which operate at 65 mK (Breon et al. 1999; Shirron et al. 2000), and important advances in cooler technology will be made for NGST. Cryogenic capillary pumped loops, which have already been tested in space, have the potential to distribute cooling power over long distances (Bugby et al. 1998).

4.3 Optics and Interferometry

The mirrors needed for far-IR/sub-mm space interferometry are similar to those needed for SAFIR, only smaller. The mirrors must: (a) be light in weight ($1 - 3$ kg m$^{-2}$), (b) have a surface roughness not exceeding ~0.5 µm rms, (c) be able to be cooled to <10 K, and (d) maintain their shape to a small fraction of a wavelength when subjected to cooling or mechanical stress. Flat mirrors, perhaps stretched membranes (Dragovan 2000), could be used for the light collecting elements of the interferometer. The additional requirements for interferometry are beamsplitters that can operate at ~4 K and over the far-IR/sub-mm wavelength range, and long-stroke cryogenic delay lines. For a 5-year SPIRIT mission the delay line would have to be able to stroke (full amplitude) at ~$10^{-2}$ Hz and survive at least $10^6$ cycles; for SPECS the ideal delay line would move 100x faster and survive a proportionately greater number of cycles. (These numbers are based on the assumption that the mirror movement is fast enough to completely sample the synthetic aperture plane in the time required to build up the typical desired sensitivity.) The delay line would have to impart minimal disturbance on the metering structure. Finally, mosaicing techniques and algorithms for wide-field interferometry will have to be developed. Research on cryogenic delay lines and beam combiners (Swain et al. 2001; Lawson et al. 2002) and wide-field imaging interferometry (Leisawitz et al. 2002; Rinehart et al. 2002) is now underway.

4.4 Large Structures and Formation Flying

A variety of architectures are possible for SPIRIT, but all of them depend on the availability of a lightweight, deployable truss structure measuring at least 30 m in length when fully expanded. Any parts of the truss that will be seen by or in thermal contact with the mirrors must be cryogenic. One possible design requires the deployed structure to be controllable in length. Another requires tracks and a mirror moving mechanism. A third design solution uses a series of mirrors along the structure to provide non-redundant baseline coverage. In all cases the boom would spin to sample different baseline orientations. Any repeating mirror movements will have to be smooth and rely on a mechanism that is robust enough to survive at least 10,000 cycles. Structures designed to meet the challenges of space-based optical interferometry have been under study for a long time for SIM (Laskin & San Martin 1989), which has far more demanding control and metrology requirements than those of SPIRIT because SIM will operate at much shorter wavelengths.

Free-flying spacecraft will be needed to accomplish imaging interferometry with maximum baseline lengths in the 1 km range. The requirement is to sample the u-v plane completely, yet avoid the need for an unaffordable amount of propellant for formation flying. It may be necessary to combine tethers with formation flying to form a long-baseline observatory that maintains symmetry while rotating. The system will have to be deployable, stable, and capable of being pointed at a variety of targets. A modeling effort is now underway, and early results suggest that tethered formation flying is feasible (Farley & Quinn 2001).

4.5 Technology Validation

> *"SAFIR, the [UVOIR from Space] panel's top-priority moderate-size mission, … will enable a distributed array in the decade 2010 to 2020…The single most important requirement is improved angular resolution. The logical build path is to develop a large, single-element (8-m class) telescope leveraging NGST technology on time scales set by NGST's pace of development. A later generation of interferometric arrays of far-infrared telecopes could then be leveraged on SIM or TPF technologies …."* – Decade Report, Panel Reports, p. 329.

Table 3 shows that there could be a rich heritage in space-validated technologies for far-IR/sub-mm interferometry by the beginning of the next decade. Ground-based laboratory or field research and testbed experiments are already underway, and more such research will be proposed to advance the technology readiness of components (e.g., detectors and array readout devices), systems (e.g., cryogenic delay line), or techniques (e.g., wide-field imaging interferometry) this decade.

**Table 3**. Technology Heritage for Long-baseline Far-IR/Sub-mm Interferometry [a]

| Technology | SIRTF 2002 | SOFIA 2004 | Astro-E2 2005 | StarLight 2006 | NGST 2009 | SIM 2009 | SPIRIT 2010 | ST-? [b] ~2010 | SAFIR 2012 |
|---|---|---|---|---|---|---|---|---|---|
| Detectors | X | * | | | | | ** | | ** |
| Coolers | X | | * | | * | | ** | | ** |
| Optics & Interferometry | X | | | *  ** | * | * | ** | | ** |
| Large structures & Formation Flying | | | | ** | * | * | X | ** | X |

NOTE: X denotes mission contributing to technology development; * denotes mission critical to success of SPIRIT (similar to technology inheritance for SAFIR); ** denotes mission critical to success of SPECS
[a] TPF will contribute substantially to the technology heritage if an interferometric solution is selected from among several concepts under consideration
[b] A hypothetical New Millennium Mission designed to validate tethered formation flying

5. Recommendations

> *"A rational coordinated program for space optical and infrared astronomy would build on the experience gained with NGST to construct SAFIR, and then ultimately, in the decade 2010 to 2020, build on the SAFIR, TPF, and SIM experience to assemble a space-based, far-infrared interferometer."* – Decade Report, p. 110.

A coordinated, intensive technology program this decade is the key to success on this timescale. The critical technology areas outlined in section 4 – detectors, cooling systems and components, large optics, interferometric techniques, cryogenic delay lines, deployable structures, and formation flying – deserve particular attention. Much of this investment will apply to SAFIR as well as far-IR/sub-mm interferometry.

A study program for far-IR/sub-mm space astronomy should be initiated as soon as possible. To ensure that the technology funds will be wisely invested it is essential to take a system-level look at the scientific, technical, and design tradeoffs. SAFIR and far-IR/sub-mm interferometry concepts could be studied together to ensure that each mission takes the best advantage of its architecture type, and to explore the possibility that overall cost savings could accrue through, for example, reuse of test facilities, hardware, design solutions, and coordinated technology validation. The study might identify presently unplanned but necessary technology demonstration experiments.

Acronyms

ALMA – Atacama Large Millimeter Array
HST – Hubble Space Telescope
NGST – Next Generation Space Telescope
ISO – Infrared Space Observatory
SAFIR – Single Aperture Far-IR Telescope
SIM – Space Interferometry Mission
SIRTF – Space Infrared Telescope Facility
SOFIA – Stratospheric Observatory for IR Astronomy
SPECS – Submillimeter Probe of the Evolution of Cosmic Structure
SPIRIT – Space Infrared Interferometric Telescope
TPF – Terrestrial Planet Finder